%% file: ScienceRun2024-DMe.tex
\makeatletter
\@ifundefined{@parse@version@dash}{%
\def\@parse@version#1{\@parse@version@0#1}
\def\@parse@version@#1/#2/#3#4#5\@nil{%
\@parse@version@dash#1-#2-#3#4\@nil}
\def\@parse@version@dash#1-#2-#3#4#5\@nil{%
  \if\relax#2\relax\else#1\fi#2#3#4 }
}{}
\makeatother
\documentclass[reprint,showkeys, superscriptaddress,aps,amsmath,amssymb,notitlepage,prl]{revtex4-2}
\usepackage[dvipsnames]{xcolor}

\usepackage{graphicx}
\usepackage{dcolumn}
\usepackage{bm}
\usepackage{multirow}
\allowdisplaybreaks
\usepackage{verbatim}
\usepackage{xspace} 
\usepackage{float}
\usepackage{booktabs}
\usepackage[colorlinks]{hyperref}
\usepackage{lineno}
\usepackage{soul}

\newcommand\myshade{80}
\colorlet{mylinkcolor}{ForestGreen}
\colorlet{mycitecolor}{Red}
\colorlet{myurlcolor}{violet}

\hypersetup{
  linkcolor  = mylinkcolor!\myshade!black,
  citecolor  = mycitecolor!\myshade!black,
  urlcolor   = myurlcolor!\myshade!black,
  colorlinks = true
}

\newcommand{\geant}{\textsc{Geant4}\xspace}

\begin{document}

\title{Probing Benchmark Models of Hidden-Sector Dark Matter with DAMIC-M}

\input{0_authorlist}

\date{\today}%

\begin{abstract}
We report on a search for sub-GeV dark matter (DM) particles interacting with electrons using the DAMIC-M prototype detector at the Modane Underground Laboratory. The data feature a significantly lower detector single $e^-$ rate (factor 50) compared to our previous search, while also accumulating a ten times larger exposure of $\sim$1.3\,kg-day. DM interactions in the skipper charge-coupled devices (CCDs) are searched for as groups of two or three adjacent pixels with a total charge between 2 and 4\,$e^-$. We find 144 candidates of 2\,$e^-$ and 1 candidate of 4\,$e^-$, where 141.5 and 0.071, respectively, are expected from background. With no evidence of a DM signal, we place stringent constraints on DM particles with masses between 1 and 1000\,MeV/$c^2$ interacting with electrons through an ultra-light or heavy mediator. For large ranges of DM masses below 1\,GeV/c$^2$, we exclude theoretically-motivated benchmark scenarios where hidden-sector particles are produced as a major component of DM in the Universe through the freeze-in or freeze-out mechanisms.
\end{abstract}


\maketitle

\textit{Introduction.---}
The existence of a non-baryonic and non-luminous form of matter~\cite{Bertone:2004pz}, known as dark matter (DM), is supported by numerous astrophysical and cosmological observations. A compelling explanation is that DM is made of hitherto unknown particles, including Weakly Interacting Massive Particles (WIMPs) with masses larger than the proton’s ($\sim 1~\rm{GeV}/c^2$). 
However, its direct detection through nuclear recoils induced in the target material~\cite{Drukier:1983gj,Goodman:1984dc,Drukier:1986tm}
 has so far eluded experimental efforts~\cite{Billard:2021uyg}.
Light (sub-GeV) DM particles \textemdash~WIMPs as well as others from a range of natural, well-motivated scenarios~\cite{Boehm:2003ha,Hooper:2008im,Pospelov:2007mp,Hooper:2008im,Chu:2011be,Knapen:2017xzo} \textemdash~are also viable candidates. 
Their discovery through nuclear recoils is challenging since the signal falls below the energy threshold for detection~\cite{Essig:2011nj}. 
On the other hand, electronic recoils produced by DM-\textit{electron} (DM-$e^-$) scattering are observable in semiconductor-based detectors with $\mathcal{O}$(1 eV) ionization thresholds~\cite{Essig:2015cda}. 
In particular, hidden-sector DM candidates that interact via a new gauge boson feebly mixed with the photon~\cite{Holdom1986,Okun1982} become experimentally accessible.

The rate of DM-$e^-$ scattering in a semiconductor is~\cite{Essig:2015cda,PhysRevD.109.115008}:
\begin{align}
\label{eq:dR_dEe}
\begin{split}
    \frac{\mathrm{d}R}{\mathrm{d}E_e} \propto \bar{\sigma}_e \int \frac{\mathrm{d}q}{q^2}\left[\int\frac{f(\mathbf{v})}{\mathrm{v}}
    \,\mathrm{d^3v}\right]F_\mathrm{DM}(q)^2\left|F_\mathrm{c}(q,E_e)\right|^2\,,
\end{split}
\end{align}
where $\bar{\sigma}_e$ is the DM-$e^-$ scattering  reference cross-section, evaluated at a momentum transfer $q = \alpha m_e$~\cite{Essig:2015cda},  $f(\mathbf{v})$ is the DM velocity distribution, $F_\mathrm{c}$ is the product of the crystal form factor and the screening factor~\cite{Essig:2015cda,Lee:2015qva,Griffin_2021,PhysRevD.109.115008}, and $E_e$ is the energy deposited as ionization. 
For hidden-sector particles, the DM form factor $F_\mathrm{DM} = (\alpha m_e/q)^n$ depends on the mass of the mediator $m_{A^\prime}$, with $n=0$ for a heavy mediator ($m_{A^\prime} \gg \alpha \, m_e$) and $n=2$ for ultra-light mediators ($m_{A^\prime} \ll \alpha \, m_e$)~\cite{Emken:2019tni}. In these models the correct DM relic abundance can be reproduced, for example, by thermal “freeze-out” of DM annihilation into Standard Model particles during the early Universe (heavy mediator) or by “freeze-in” of DM produced by the annihilation of Standard Model particles (ultra-light mediator)~\cite{Hall:2009bx}. 
Although benchmark theoretical predictions for these scenarios~\cite{Chu:2011be,Essig:2015cda,Izaguirre:2015yja,USCosmicVisions2017,PhysRevD.99.115009, PhysRevD.105.119901, PhysRevD.110.L031702} have been the target of many experimental efforts
~\cite{DAMIC-M:2023gxo,
PhysRevLett.132.101006,PhysRevLett.130.101002,PhysRevLett.130.261001,PhysRevLett.134.011804,PhysRevD.111.012006,aprile2024searchlightdarkmatter,xenon1t2019,DAMIC:2019dcn,EDELWEISS:2020fxc,SuperCDMS:2018mne,cdex10}, they have been only marginally scrutinized due to lack of sensitivity.


The DAMIC-M (Dark Matter in CCDs at Modane) experiment~\cite{Privitera:2024} employs skipper charge-coupled devices (CCDs) with unprecedented sensitivity to DM-induced electronic recoils, thanks to their sub-electron readout noise and extremely low level of dark current. We previously reported exclusion limits on sub-GeV DM particles interacting with electrons~\cite{DAMIC-M:2023gxo,PhysRevLett.132.101006} from prototype CCDs in the Low Background Chamber (LBC)~\cite{arnquist2024damicmlowbackgroundchamber} at the Modane Underground Laboratory (LSM \footnote{Laboratoire Souterrain de Modane}). 
Here we present a much more sensitive search, which probes for the first time benchmark predictions for hidden-sector DM over a wide range of DM masses.

\textit{Setup and data.---} 
Two CCD modules that implement the final DAMIC-M design are installed in the LBC. 
Each module includes four high-resistivity ($>$10\,k$\Omega$\,cm), $n$-type silicon CCDs~\cite{Holland:2002zz,Holland:2003zz,Holland:2009zz} glued to a silicon pitch adapter with traces for clocks, biases and video signals. Each CCD has $6144\times 1536$ pixels (pixel area $15\times 15\,\mu$m$^2$) and is $670\,\mu$m thick, for a mass of $\sim$$3.3$\,g. In the CCD module, only one amplifier out of four is read out from each CCD. The detectors are mounted in a high-purity copper box, which also shields infrared radiation. The CCDs are kept at a temperature of $\sim130$~\,K inside the LBC vacuum cryostat at $\sim$$5\times 10^{-6}$\,mbar. A lead shield of $\sim$$7.5$\,cm thickness, with the innermost 2\,cm of ancient origin~\cite{ALESSANDRELLO1991106}, encloses the copper box inside the cryostat. An additional 15\,cm of lead and 20\,cm of high density polyethylene in a movable structure surround the cryostat to shield from environmental radiation. Recent changes in the LBC setup~\cite{arnquist2024damicmlowbackgroundchamber} include the new CCD modules, better light-tightness of the box, and the final DAMIC-M CCD controller and front-end electronics. These changes result in improvements of the dark current, and of the single electron resolution with respect to Ref.~\cite{DAMIC-M:2023gxo}.

Charge carriers produced by ionizing radiation in the silicon bulk of the CCD drift towards the $x$-$y$ plane where pixels are located. Due to thermal diffusion while drifting, the charge collected by the pixels presents a spatial variance, $\mathrm{\sigma^{2}_{xy}}$, which correlates with the depth of the interaction~\cite{PhysRevD.94.082006}. At the applied substrate bias voltage of 45\,V, $\mathrm{\sigma_{xy}}\sim1.2$~pixel for the maximum drift length of 670\,$\rm{\mu m}$. In the readout process, the pixel charge is moved across the CCD array by appropriate voltage clocking, also allowing for the charge of several pixels to be binned before being read out. 
The charge is read out serially by one amplifier at the end of the serial register. The skipper amplifier~\cite{skipper,Chandler1990zz,Tiffenberg:2017aac} allows for multiple non-destructive charge measurements (NDCMs) of the pixel charge.  Sub-electron resolution 
is achieved by averaging a sufficient number of NDCMs. The pixel charge resolution, $\sigma_{ch}$, is $\sim 0.16\,e^-$ for the data reported in this Letter.

For this analysis 84\,days of data is used, taken between October 2024 and January 2025. To improve electrical grounding, we left the detector’s movable shield slightly open during this period \footnote{The readout noise of skipper CCDs is sensitive to electrical grounding, which was not fully optimized at that time. After an intervention following the completion of the data taking period, we achieved $\sigma_{ch} < 0.16\,e^-$ with the shielding fully closed.}, which resulted in an increased radiogenic background from 7~\cite{arnquist2024damicmlowbackgroundchamber} to $\sim$$15$\,events/keV/kg/day (dru).
Since a 100-pixel binning in the vertical direction ({\it i.e.} rows) is used in the synchronous read out of the eight CCDs, in subsequent text the term ``pixel'' refers to the binned pixel. The CCD pixels are read out continuously while simultaneously accumulating dark matter exposure, with data recorded in images of 6300 columns and 16 rows~\footnote{We read out more columns than the 6144 physical ones of the CCD. These additional pixels, which have much lower exposure, provide useful information on data quality and dark current in the serial register.}. The corresponding exposure of each CCD pixel is approximately equal to the image readout time of 1668 s. For each pixel in each image we also acquire the variance $\sigma^2_{\rm{NDCM}}$ of its NDCMs.   

\textit{Pixel selection.---}
Selection criteria are established with the first seven days of data (labelled “D1”), and subsequently applied to the remaining data set through a blind analysis (“D2”).
One CCD (2-A, where the number 2 identifies the module and the letter A the CCD within the module) has an anomaly at the end of the serial register that results in an excess of pixels with charge  $>$1\,$e^-$, and is excluded from the analysis. A second CCD (2-C) with much larger noise is also excluded. A total of six CCDs is thus retained for the analysis.
Image data reduction and pixel selection criteria are briefly summarized here, and detailed in the End Matter.
For each CCD, images are first stitched together in a single image, which is then calibrated from the fitted position of the 0\,$e^-$ and 1\,$e^-$ peaks in the pixel-value distribution. 
Defects in the CCDs, appearing in the images as ``hot" columns~\cite{2001sccd.book.....J}, are identified by their larger rate of 1$e^{-}$ pixels, and rejected. A high dark current region of CCD 1-C following a prominent hot column (5315) is also excluded from the analysis. Then, contiguous pixels with charge are joined into clusters. Those with total charge $\ge$\,6$e^-$ are excluded from further analysis since their expected contribution to a DM-$e^-$ signal is negligible.
To reject crosstalk and correlated noise, pixels belonging to a cluster are masked in all CCDs of the same module, as well as pixels with high $\sigma^2_{\rm{NDCM}}$ and pixels whose values are positively or negatively correlated in all CCDs. To eliminate trailing charge released from traps in the CCD active area or in the serial register, we mask 100 pixels above any pixel with charge $>100\,e^-$ as well as all pixels in its row. Lastly, we reject rows and columns having a large number of pixels with charge $\ge 1\,e^-$~\footnote[51]{We define the charge q of a pixel to be $ \ge n\,e^-$ when q $\ge (n-1)+3.75\sigma_{ch}$} or with more than one pixel with charge $\ge2\,e^-$ (unless these pixels are consecutive along a row). These criteria select $95\%$ of the pixels in the CCD active area, resulting in an integrated exposure of 0.139\,kg-days for D1 and 1.257\,kg-days for D2. The measured single $e^-$ rate in D2 is in the range $2.4-3.1\cdot 10^{-4}\,e^-$/pixel/image, equivalent to $350-460\,e^-$/g/day. Compared to our previous results~\cite{DAMIC-M:2023gxo}, we achieve a tenfold increase in exposure and a factor of 50 reduction in detector single $e^-$ rate.
 
\textit{Candidate selection.---}
The pixel selection criteria applied to the D1 data set reveal an excess of $2\,e^-$ pixels (36, with 11.6 expected from a Poissonian single $e^-$ rate; there are no pixels with charge of 3 or $4\,e^-$). While its origin is still to be fully understood, we do observe a much more significant excess in CCD 2-A, which suggests that it may be related to the serial register or readout stage. A DM search based on isolated pixels would be limited by this background. On the other hand, a DM interaction in the silicon bulk of the CCD typically results in charge spread over more than one pixel due to diffusion. We thus exclude isolated pixels, and search for DM candidates as patterns of two or three consecutive pixels. Given the 100-pixel binning in the vertical direction, only patterns found in rows are considered. Specifically, we search for the following patterns and, when appropriate, their permutations: $\{11\}$, $\{21\}$, $\{111\}$, $\{31\}$, $\{22\}$, and $\{211\}$, where the number corresponds to the pixel charge in $e^-$.   
Consecutive pixels are classified as a pattern $\rm{\{mn\}}$ or $\rm{\{mnl\}}$ based on the value of a pattern identification variable. This variable quantifies the probability to obtain the observed pixel values, assuming charges of m, n, l in consecutive pixels, taking into account the charge resolution $\sigma_\mathrm{ch}$ (further details on pattern identification are given in the End Matter).

In the D1 data set we find 22 candidates for pattern $\{11\}$, consistent with the expectation of 21.5 from random coincidences of two consecutive pixels with 1\,$e^-$, and zero candidates for all other patterns.  These results provided confidence on the selection procedure before its application to the blinded data set. 

\begin{figure}[t!]
    \centering
	\hspace{-0.2cm}
    \vspace{-0.6cm}
        \includegraphics[width=0.48\textwidth, trim={1cm 1cm 1cm 1cm}, clip]{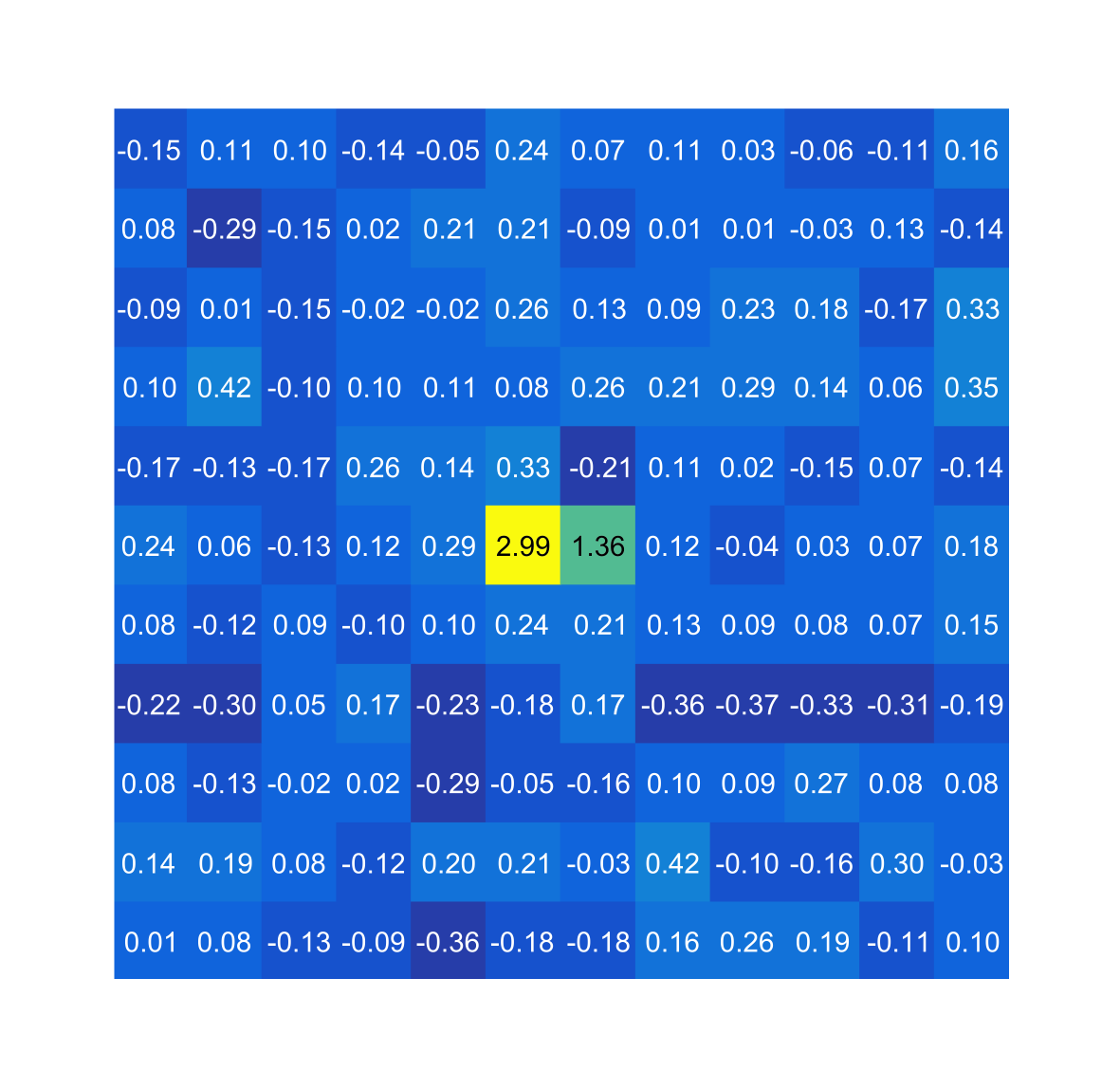}
        \caption{A portion of the CCD image with the $\{31\}$ candidate pattern, with the measured charge in $e^-$ reported on the pixels. The pattern is found in column 2103 of CCD 1-D.}
    \vspace{-0.6cm}
    \label{fig:31candidate} 
\end{figure}

After unblinding, in the D2 data set we find 144 candidates for pattern $\{11\}$, one candidate for pattern $\{31\}$, and zero candidates for all other patterns. A careful inspection of the $\{31\}$ pattern candidate, shown in Fig.~\ref{fig:31candidate}, confirms its classification as a DM candidate. 

The excess of high-multiplicity isolated pixels in D1 is confirmed in the D2 data set, with 184 pixels with $2\,e^-$, 17 with $3\,e^-$ and 1 with $4\,e^-$ to be compared with 70.2, $7\cdot 10^{-3}$, and $3\cdot 10^{-7}$ expected from a Poissonian single $e^-$ rate, respectively. 

\textit{Backgrounds.---}
Random coincidences of consecutive pixels with charge originating from background single $e^-$ may produce candidate patterns. In addition, we expect only 3.7 patterns $\{11\}$  from accidental noise in D2, where a 1\,$e^-$ pixel is adjacent to an empty pixel with an upward fluctuation of readout noise.

The corresponding number of background events, $B^\mathrm{rc}_p$, for a given pattern $p$ is determined from the measured rates of 1, 2, and 3\,$e^-$ pixels and from the pattern identification efficiency evaluated with toy Monte Carlo (MC) simulations. To account for small differences in the single $e^-$ rate as a function of column, expectations are calculated in consecutive bands of 10 columns, and then summed to obtain the total number of candidates expected in a given CCD.

Radioactive decays in the detector and surrounding material are another source of background. We use a detailed \geant~\cite{geant4} simulation of the apparatus to obtain the energy $E_e$ deposited in the CCDs by these radiogenic and cosmogenic backgrounds. The simulation includes realistic amounts of radioactive contaminants as determined by radioassay measurements and
bookkeeping of cosmogenic activation time of materials~\cite{arnquist2024damicmlowbackgroundchamber}. 
Then, $E_e$ is converted to a number $q_e$ of electron-hole pairs with 
the charge yield model in Ref.~\cite{PhysRevD.102.063026}. The probability that a given charge deposit is found as a pattern $p$, $P_{q_e\rightarrow p}$, is obtained through a MC simulation. 
Charge is diffused from the location of the deposit on the pixel array with $\mathrm{\sigma^2_{xy}}$ obtained from a surface laboratory calibration with cosmic muons~\cite{PhysRevD.105.062003,DAMIC-M:2023gxo}\footnote{$\mathrm{\sigma^2_{xy}(z)}=-a\ln|1-b\mathrm{z}|\cdot(\alpha + \beta E_e)^2$, with parameters $a = 803\,\mu\rm{m}^2$, $b = 6.5 \times 10^{-4}\,\mu\rm{m^{-1}}$, $\alpha=0.83$, and $\beta= 0.0112\,\rm{keV^{-1}}$.}. Then, 100-pixel vertical binning is applied to the simulated pixel array to match the data; the single $e^-$ background is added to the binned array; the pixel charge is fluctuated according to its resolution $\sigma_{ch}$; and the candidate selection is applied to identify patterns.

Limited by the low statistics and the small energy region, we estimate the background in the region of interest by scaling the event rate at higher energy. The number of background events from radioactive decays, $B^{\mathrm{rad}}_p$, for a given pattern $p$ is then given by:
\begin{equation}
    B^{\mathrm{rad}}_p = \tau_p N^{\mathrm{rad}}_{\mathrm{obs}},
    \label{eqn:Brad}
\end{equation}
where the scaling factor $\tau_p$ is obtained from the simulations described above, and $N^{\mathrm{rad}}_{\mathrm{obs}}=98$ is the number of clusters between 2.5 and 7.5 keV in the D2 data set. This corresponds to $\sim 15$\,dru and is consistent with expectations from a \geant\ MC simulation of the apparatus with the slightly open shield.

The expected number of background events for each pattern $p$ is reported in Table~\ref{tab:D2results}, together with the number of candidates $D_p$.
\begin{table}[ht]
\centering 
\begin{tabular}{| c | c | c  | c |} 
\hline
& \multicolumn{3}{c|}{Pattern $p$}\\ [0.5ex]
\hline
 & ~~$\{11\}$~~ & ~~$\{21\}$~~ & ~~$\{111\}$~~  \\ [0.5ex] 
 \hline
  $D_{p}$ & 144 & 0 & 0  \\ 
\hline
  $B^\mathrm{rc}_{p}$ & 141.4 & $0.111$ & $0.042$ \\
  \hline
  $B^{\mathrm{rad}}_p$ & 0.039& 0.039 & 0.016 \\
  \hline\hline
  & ~~$\{31\}$~~ & ~~$\{22\}$~~ & ~~$\{211\}$\\[0.5ex] 
  \hline
  $D_{p}$  & 1 & 0 &  0  \\ 
\hline
  $B^\mathrm{rc}_{p}$ & 0.019 &  $2.5\cdot10^{-5}$ & $5.8\cdot10^{-5}$\\
  \hline
  $B^{\mathrm{rad}}_p$ & 0.052 & 0.011 & 0.035\\
\hline 
\end{tabular}
\caption{The number of candidates $D_p$ in the D2 data set, and the number expected from backgrounds due to random coincidences, $B^\mathrm{rc}_{p}$, and to radioactive decays, $B^{\mathrm{rad}}_p$.} 
\label{tab:D2results} 
\end{table}
Notably, the number of $\{11\}$ pattern candidates found in the data is in excellent agreement with the expected background. From Table~\ref{tab:D2results}, the probability to have one or more background events in any of the patterns with 3 or more electrons is nearly 28\%, and to have one or more in the $\{31\}$ pattern is 7\%.  Hence the observed event in the $\{31\}$ pattern topology is compatible with the background expectations.

 It is worth mentioning that the low-energy excess observed by DAMIC at SNOLAB~\cite{PhysRevD.109.062007} would result in $\mathcal{O}(1)$ event in the search presented here. A dedicated analysis to verify the excess with clusters of charge $\ge6\,e^-$ (which are excluded from this analysis, and blinded to 2.5\,keV)
will be the subject of a future study.

\textit{Results.---}
To constrain a DM signal, we use the profile likelihood ratio analysis~\cite{Cowan:2010js} as recommended in Ref.~\cite{PhystatDM}. The DM signal model $S_p$ for a given pattern $p$ is:
\begin{equation}
    S_p = \sum_{q_e} N_{q_e} (\bar{\sigma}_e,m_\chi)P_{q_e\rightarrow p}\,,
\end{equation}
where $N_{q_e}$ is the number of DM interactions resulting in $q_e$ ionization charges, and $m_\chi$ is the mass of the DM particle. 
To calculate $N_{q_e}$ we multiply the DM-$e^-$ interaction rate in the detector with the integrated exposure of D2. 
For the interaction rate we use Eq.~\eqref{eq:dR_dEe} with $\texttt{QCDark}$~\cite{PhysRevD.109.115008} for the crystal form factor and analytic screening,  
and a standard Maxwell-Boltzmann DM velocity distribution with parameters of the DM density profile in the galactic halo as recommended in Ref.~\cite{PhystatDM}. The conversion of $E_e$ to electron-hole pairs and the procedure to derive $P_{q_e\rightarrow p}$ is the same previously described for the $B^{\mathrm{rad}}_p$ background, with the MC charge injected uniformly across the sensitive volume of the CCD as expected from DM interactions. 
With this procedure the efficiency for a (2, 3, 4, 5)\,$e^-$ deposit to be found as a pattern is $\sim$(38, 65, 79, 86)\%, respectively. 

The background model $B_p$ is given by:
\begin{equation}
B_p = B^\mathrm{rc}_p + \theta B^{\mathrm{rad}}_p,
\end{equation}
where $B^{\mathrm{rad}}_p$ is defined in Eq.~\eqref{eqn:Brad} and $\theta$ is a nuisance parameter constrained by the inclusion of $N^{\mathrm{rad}}_{\mathrm{obs}}$ in the likelihood as a control measurement~\cite{Cowan:2010js}. 

Following the blind analysis prescription, only the D2 data set is used in the following. We first evaluate the significance of a potential DM signal by testing $H_0$, the background-only null hypothesis ($\bar{\sigma}_e=0$). With a minimum $p_0$-value of 0.24 (0.10) 
for an ultra-light (heavy) mediator, we do not reject $H_0$. Consequently, we derive exclusion limits using the two-sided profile likelihood ratio test statistic, $\tilde{t}_\mu = -2\log\lambda(\bar{\sigma}_e)$, where $\lambda(\bar{\sigma}_e)$ is the profile likelihood ratio at each DM mass. We do not assume asymptotic approximations for $\tilde{t}_\mu$, and instead construct the test statistic distributions with MC techniques. 

\begin{figure*}
    \centering
	\hspace{-0.2cm}
        \includegraphics[width=0.48\textwidth]{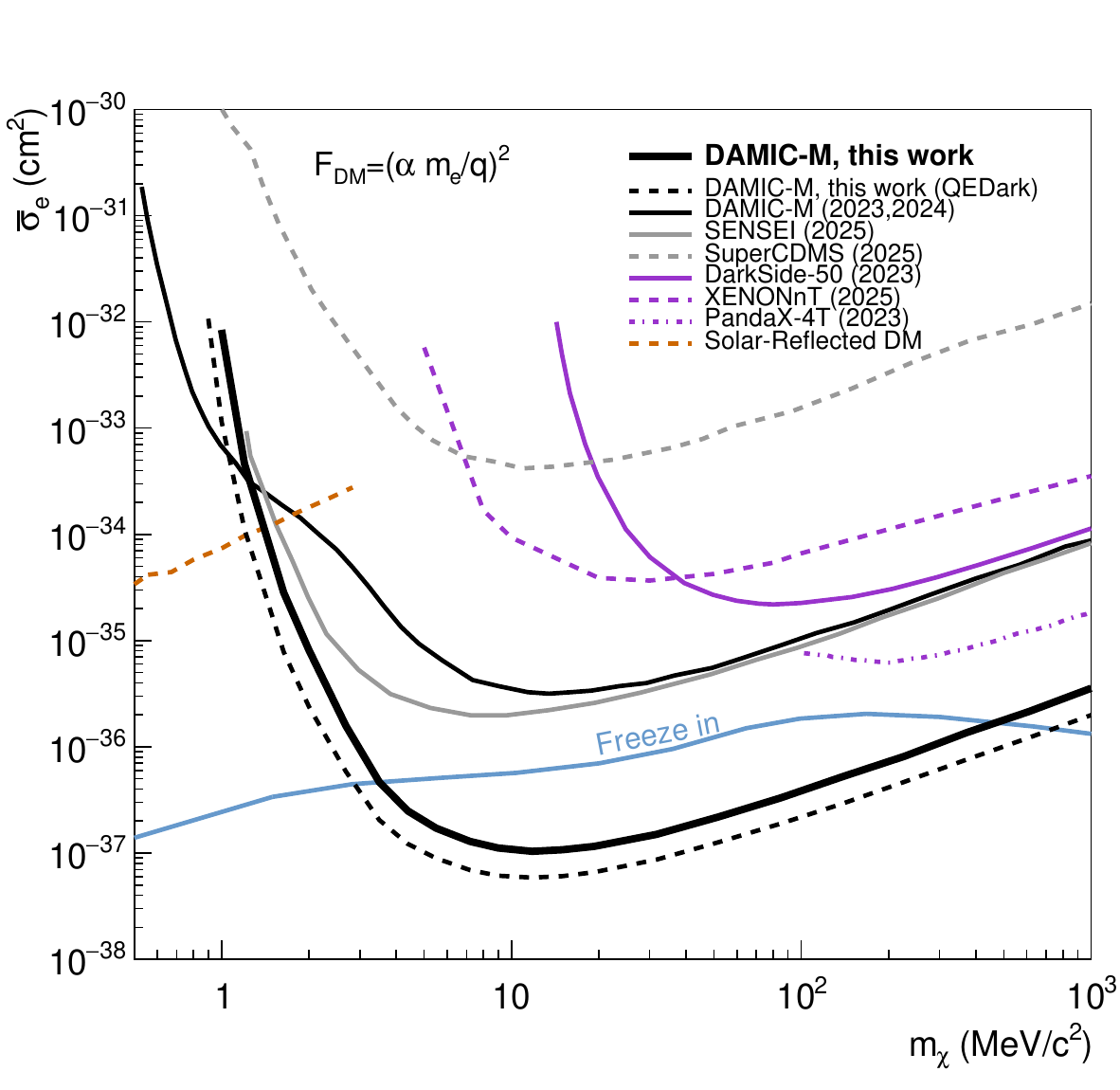}
        \hspace{0.6cm}
   	\includegraphics[width=0.48\textwidth]{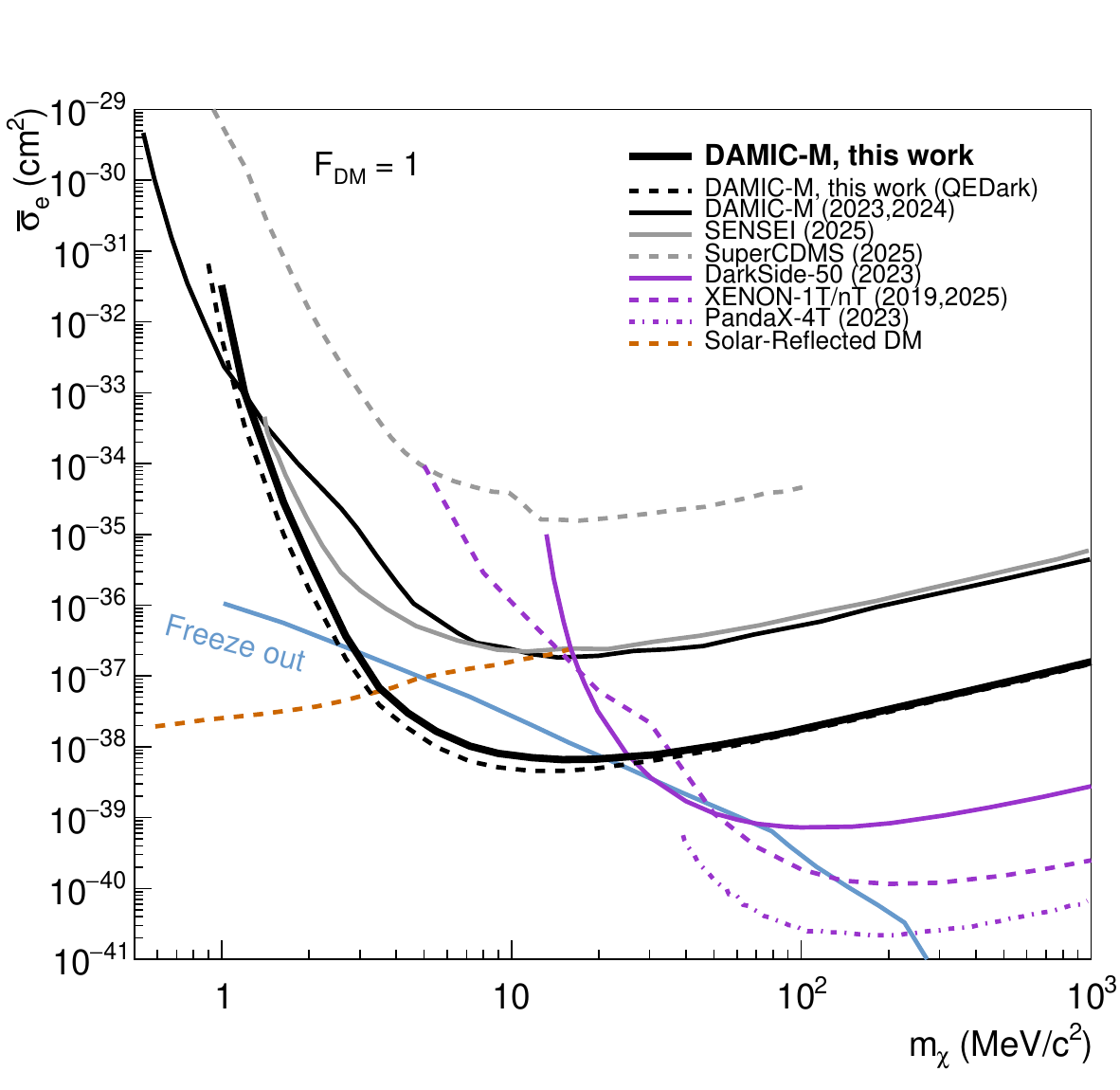}
    \caption{
    DAMIC-M 90\% C.L. upper limits (solid thick black) on DM-electron interactions through an ultra-light (\textbf{left}) and heavy (\textbf{right}) mediator. Limits based on the \texttt{QEDark} calculations (solid dashed black) are included for comparison with previous results. The previous DAMIC-M results~\cite{DAMIC-M:2023gxo,PhysRevLett.132.101006} are also shown together with limits from 
    DarkSide-50\,\cite{PhysRevLett.130.101002}, 
    PandaX-4T~\cite{PhysRevLett.130.261001},
    SENSEI~\cite{PhysRevLett.134.011804}, 
    SuperCDMS~\cite{PhysRevD.111.012006}, XENONnT~\cite{aprile2024searchlightdarkmatter}, and XENON1T~\cite{xenon1t2019} (also reinterpreted in Ref.~\cite{Emken_2024} in terms of ``solar-reflected DM").
    Theoretical expectations assuming a DM relic abundance from freeze-in and freeze-out mechanisms are also shown in light blue~\cite{USCosmicVisions2017} (the freeze-out benchmark is valid for the specific scenario of complex scalar DM with $m_{A'} \gg m_\chi$). At very large cross-sections, this search is no longer sensitive to DM particles as they are slowed or stopped in the overburden above the detector. For the ultra-light mediator, this cross-section ``ceiling" lies above $\bar{\sigma}_e = 10^{-27}\,\mathrm{cm}^2$ over the mass range of interest. For the heavy mediator, this ceiling ranges from $\bar{\sigma}_e = 2\times 10^{-27}\,\mathrm{cm}^2$ at 1 MeV/$c^2$ to $\bar{\sigma}_e = 7\times 10^{-36}\,\mathrm{cm}^2$ at 1 GeV/$c^2$ \footnote{We calculate the effects of Earth-shielding using the verne code~\cite{Kavanagh:2017cru}, assuming that the mediator is a dark photon and that scattering off nuclei dominates the shielding effect~\cite{Emken:2019tni}.}. We note that cross-sections above this ceiling are already excluded by detectors operated on the surface and by other probes~\cite{Emken:2019tni}.
    }
\label{fig:upperlimits}
 \end{figure*}

The 90\% C.L. exclusion upper limits for dark matter interacting via ultra-light (left) and heavy (right) mediators are shown in Fig.~\ref{fig:upperlimits}. Also shown are previous results from DAMIC-M and other direct detection experiments~\footnote{For better legibility of the figures we include only the most recent results. Other exclusion limits can be found in Refs.~\cite{xenon10_2017, DAMIC:2019dcn, EDELWEISS:2020fxc, SuperCDMS:2018mne, cdex10}. Other solar-reflected DM limits can be found in Refs.~\cite{PhysRevD.104.103026, PhysRevD.105.063020, PhysRevLett.132.171001, aprile2024searchlightdarkmatter}.}
  along with benchmark theoretical predictions for cross-sections where the freeze-in and freeze-out mechanisms lead to the observed DM abundance in our Universe today. For comparison with previous results~\cite{DAMIC-M:2023gxo,PhysRevLett.132.101006,PhysRevLett.134.011804,PhysRevD.111.012006} we include in Fig.~\ref{fig:upperlimits} limits using $\texttt{QEDark}$~\cite{Essig:2015cda} for the crystal form factor in Eq.~\eqref{eq:dR_dEe} and no screening (theoretical calculations of the rate are further discussed in the End Matter). 
   
Our exclusion limits fall within the expected 68.3\% sensitivity band over the entire
mass range for an ultra-light mediator, and up to $m_{\chi}\sim 30\,\rm{MeV}/c^2$ for a heavy mediator. The heavy mediator limit is always within the expected 80\% sensitivity band. Cross-checks of the analysis are included in the End Matter.

\textit{Conclusions and Outlook.---}
With this search DAMIC-M establishes the most stringent limits to date on DM particles interacting with electrons over a large range of DM masses between 1 and 1000\,MeV/$c^2$. 

Most notably, these limits make it possible for the first time to rule out that DM particles interacting feebly with electrons via an ultra-light mediator could have been produced as a dominant component of DM in the Universe by the freeze-in mechanism, for DM masses between 3.5 and 490\,MeV/c$^2$. In addition, they also rule out that \textit{any} fraction of the DM abundance is made up of complex scalar particles with masses between 2.9 and 21.5\,MeV/c$^2$ that interact with electrons via a mediator heavier than the DM mass ($m_{A'} \gg m_\chi$). These would decouple via freeze-out in the early Universe, with cross sections \textit{below} the freeze-out benchmark exceeding the observed DM density. Taking into account all constraints of Fig.~\ref{fig:upperlimits} (right panel), only complex scalar DM particles with a mass around 25\,MeV/c$^2$ remain viable in this scenario. 

Note that the benchmark hidden-sector predictions are model-dependent~\cite{Essig:2015cda} and smaller cross sections are still possible in certain scenarios.
DAMIC-M, which plans to collect a kg-year exposure, will probe the freeze-in scenario beyond the benchmark model up to GeV DM masses as well as other hidden-sector scenarios with heavy mediators that match the relic density observed today with smaller DM-$e^-$ cross sections than that of the freeze-out scenario with scalar particles~\cite{Essig:2015cda,Izaguirre:2015yja}.

Constraints on other processes beyond the DM-$e^-$ scattering of Eq.~\eqref{eq:dR_dEe} that produce a charge ionization signal are reported in the End Matter.

These results demonstrate the excellent performance of the DAMIC-M prototype detector. The CCDs feature minimal defects and a very low dark current, allowing for highly efficient pixel selection. 
With only a few percent change of the single $e^-$ rate over the three months of data collection, the detector is remarkably stable enabling a search for daily modulation in the DM signal~\cite{PhysRevLett.132.101006} to be reported in a forthcoming publication. 
We expect to improve these results by orders of magnitude with the full DAMIC-M apparatus, soon to be installed at the LSM, which will operate $\sim$200 CCDs under strict control of radiogenic and cosmogenic backgrounds.

\textit{Acknowledgments.---}
We would like to thank the Modane Underground Laboratory and its staff for support through underground space and logistical and technical services. LSM operations are supported by the CNRS, with underground access facilitated by the Soci\'et\'e Fran\c caise du Tunnel Routier du Fr\'ejus. The DAMIC-M project has received funding from the European Research Council (ERC) under the European Union’s Horizon 2020 research and innovation program Grant Agreement No. 788137, and from the NSF through Grant No. NSF PHY-1812654. The work at the University of Chicago and University of Washington was supported through Grant No. NSF PHY-2413013. This work was supported by the Kavli Institute for Cosmological Physics at the University of Chicago through an endowment from the Kavli Foundation. We also thank the Krieger School of Arts \& Sciences at Johns Hopkins University for its contributions to the DAMIC-M experiment. The IFCA was supported by Project  DMPHENO 2 LAB (PID2022-139494NB-I00) financed by MCIN/AEI/ 10.13039/501100011033/FEDER, EU. N. C.-M. acknowledges funding from the Ramón y Cajal Grant RYC2022-038402-I, financed by MCIN/AEI/10.13039/501100011033 and the FSE +. The University of Z\"{u}rich was supported by the Swiss National Science Foundation. We thank the Boulby Underground Laboratory, SNOLAB, Canfranc Underground Laboratory, Washington Nanofabrication Facility and their staff for support through underground space, logistical and technical services. The CCD development work at Lawrence Berkeley National Laboratory MicroSystems Lab was supported in part by the Director, Office of Science, of the U.S. Department of Energy under Contract No. DE-AC02-05CH11231. We thank Teledyne DALSA Semiconductor for CCD fabrication.



\bibliography{ScienceRun2024-DMe}

\clearpage
\newpage
\section*{End Matter}

\textit{Image Data Reduction and Pixel Selection.---}
A CCD image consists of an array of values, in ADUs (analog-to-digital units), proportional to the collected charge in each pixel. We first perform a pedestal subtraction row-by-row by subtracting the lowest peak in the row's pixel-value distribution from the value of each pixel in the row. An analogous procedure is then applied column-by-column. This effectively eliminates transient baseline effects across individual CCDs. The pedestal-subtracted pixel-value distribution presents a prominent peak at zero, corresponding to zero charge, followed by a peak corresponding to pixels with 1\,$e^-$ charge. The calibration constant in ADU/$e^-$ is obtained from the distance between the 0\,$e^-$ and 1\,$e^-$ peak. 

The following selection criteria are then applied in sequence to the calibrated images: \\
\textit{Step 1} - For each column, we calculate the number of pixels with charge $\rm{q}\ge1\,e^-$~\footnotemark[51], $N^{\rm{col}}_{\rm{q}\geq1e^-}$. Columns are marked as hot and rejected when $N^{\rm{col}}_{\rm{q}\geq1e^-}> \rm{MED_{col}} + 11\cdot MAD_{col}$, where $\rm{MED_{col}}$($\rm{MAD_{col}}$) is the median(median absolute deviation) of $N^{\rm{col}}_{\rm{q}\geq1e^-}$ across the CCD.
The 5 columns before and after a hot one are also excluded from the analysis. \\
\textit{Step 2} - We find clusters of contiguous pixels with charge (q~$>3\,\sigma_{ch}$), with at least one having q~$>5e^- + 3\,\sigma_{ch}$. We further collect 2(1) pixels immediately adjacent in the horizontal(vertical) direction to cluster pixels, independently of their charge.  
Clusters are masked in the CCD where they are found as well as in the other CCDs of the same module, to eliminate possible crosstalk effects.\\
\textit{Step 3} - For any high-charge pixel with $\rm{q}>100\,e^-$, we mask the 100 pixels above and its entire row. This cut rejects trailing charge released from traps in the CCD active area or in the serial register and is only applied to the CCD where the pixel is found. The high-charge pixel is also masked in all the other CCDs of the same module. \\
\textit{Step 4} - Pixels are masked when variance $\sigma^2_{\rm{NDCM}}$ deviates from the mean by more than 6 standard deviations. Correlated noise results in pixels with high positive or negative values across all CCDs of the same module, which we reject with the
following variable:
\begin{equation}
\rm{p_{corr} = -log(\prod_{L} F_{q_{i,j,L}}(0))},
\label{eqn:pcorr}
\end{equation}
where $\rm{q_{i,j,L}}$ is the value of pixel in column $\rm{i}$, row $\rm{j}$, and CCD $\rm{L}$ (excluding CCD 2-C), and $\rm{F_{q_{i,j,L}}(m)}$ is the cumulative distribution function of a Gaussian with mean m and standard deviation $\sigma_{ch}$. We reject pixels with positive (negative) correlated values by requiring $\rm{p_{corr}}>~0.01$~($<20$). \\
\textit{Step 5} - A more stringent requirement on charge multiplicity is then applied by rejecting columns/rows with $N^{\rm{col/row}}_{\rm{q}\geq1e^-}> \rm{MED_{col/row}} + 4.45\cdot MAD_{col/row}$, with MED and MAD now calculated on pixels remaining after {\it Step~4}. We also mask columns and rows with ${N^{\rm{col/row}}_{\rm{q}\geq2e^-}>1}$, unless these pixels are consecutive along a row. \\
\textit{Step 6} - Lastly, since our search is based on patterns of consecutive pixels we reject isolated columns, \textit{i.e.} having both adjacent columns previously masked. 

These selection criteria, established with D1 and then applied to D2 after unblinding, have similar performance across data sets and CCDs. 
The fraction of pixels in the CCD active area that survive each of the selection steps, $f_{sel}$, is reported in Table \ref{tab:eff} for two CCDs, with similar values for the other CCDs~\footnote{For CCD 1-C, $f_{sel}=0.8316$ after $Step~6$ because of columns $>5316$ being masked; when this is taken into account, the fraction of surviving pixels is still around 95\%.}. Our selection keeps $\sim95\%$ of the CCD active area.

\begin{table}[ht]
\centering 
\begin{tabular}{| c | c | c |} 
\hline
& \multicolumn{2}{c|}{$f_{sel}$}\\ [0.5ex]
\hline
~~ \textit{Step} ~~ & ~~CCD 1-D~~ & ~~ CCD 2-D~~ \\ [0.5ex] 
 \hline
\textit{1+2} & 0.9862 & 0.9820   \\ 
\textit{3} & 0.9747 & 0.9724   \\ 
\textit{4} & 0.9746 & 0.9722  \\ 
\textit{5} & 0.9431 & 0.9492  \\ 
\textit{6} & 0.9414 & 0.9474  \\ 
\hline
\end{tabular}
\caption{The fraction of pixels in the CCD active area that survive each of the selection steps for two of the six CCDs in the D2 data set.
} 
\label{tab:eff} 
\end{table}

\textit{Pattern Identification.---}
For two consecutive pixels (i,j) and (i+1,j) (i=column,j=row), we define the pattern identification variable: 
\begin{equation}
\rm{p_{mn} = -log(F_{q_{i,j}}(m)\ F_{q_{i+1,j}}(n))},
\label{eqn:pmn}
\end{equation}
where $\rm{F_{q_{i,j}}(m)}$ is a parameter of Eq.~\ref{eqn:pcorr}.
 The pattern identification variable 
 $\rm{p_{mnl}}$ for three consecutive pixels is obtained by including the multiplicative term $\rm{F_{q_{i+2,j}}(l)}$ in Eq.~\eqref{eqn:pmn}. Two consecutive pixels are selected and classified as a $\rm{\{mn\}}$ pattern if $\rm{p_{mn}<4}$, where $\rm{\{mn\}}$ is the pattern with the highest total charge satisfying the requirement.  If $\rm{p_{mnl}<5.5}$ when adding a third consecutive pixel, the pattern is classified as $\rm{\{mnl\}}$. The values chosen for the $\rm{p_{mn}}$ and $\rm{p_{mnl}}$ thresholds guarantee an $\sim90\%$ efficiency to correctly classify a pattern, {i.e.} m, n and l of the pattern match the true charge of the corresponding pixel. For consecutive pixels with true charge $5\,e^-$, the efficiency to be classified as any of the considered patterns is $>90$\%. 
 
 In addition, we require the charge of pixels in the rows above and below the pattern to be $<1\,e^-$; the charge of the pattern pixels in the other CCDs to be $<1e^-$; and no other pattern or pixel with charge $\ge\,2e^-$ to be present in the columns of the candidate pattern. 

\begin{figure*}[t!]
    \centering
    \hspace{-0.2cm}
   	\includegraphics[width=0.48\textwidth]{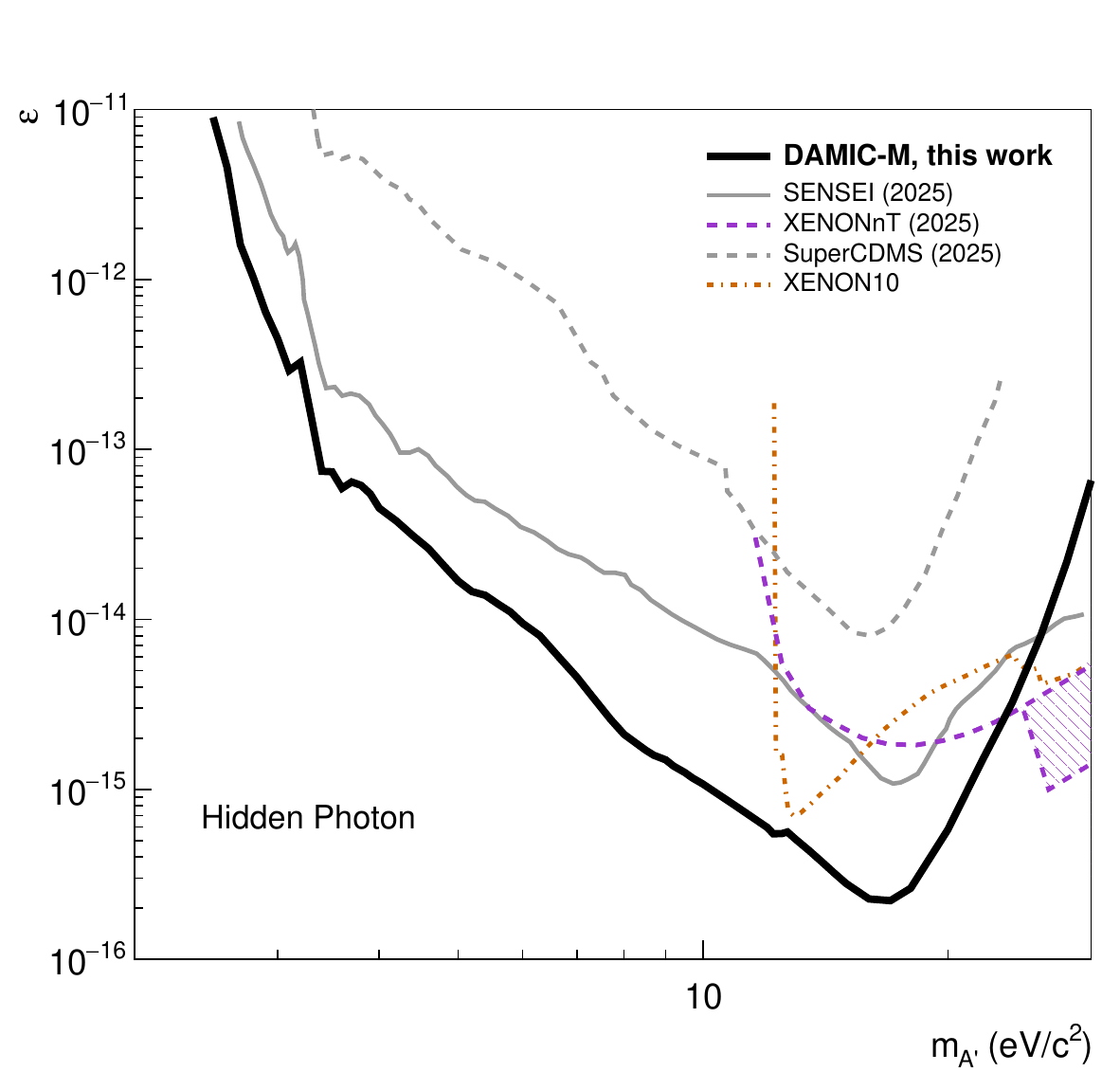}
    \hspace{0.6cm}
    \includegraphics[width=0.48\textwidth]{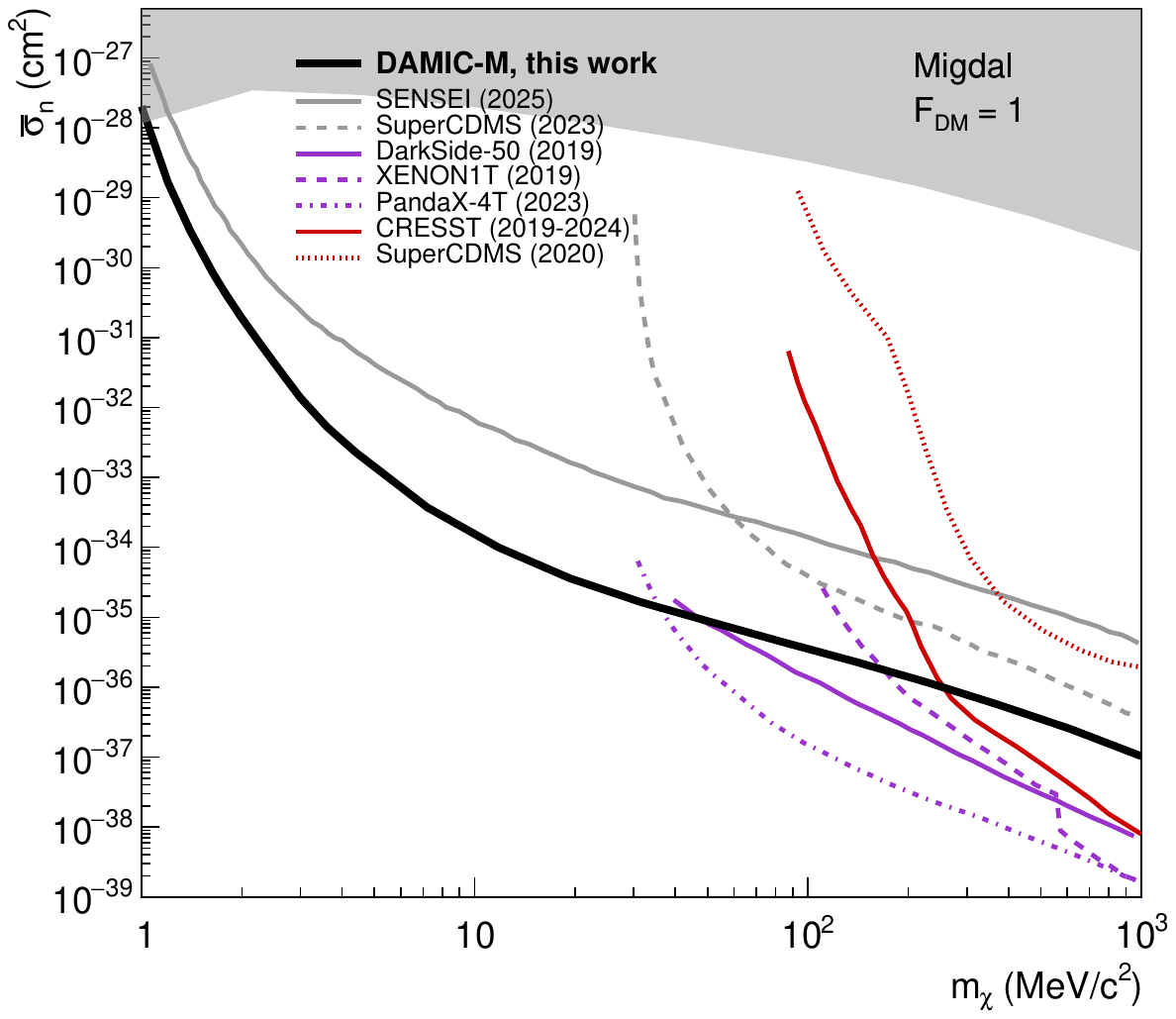}  

    \caption{(\textbf{left}) DAMIC-M 90\% C.L. upper limits (solid thick black) on the kinetic-mixing parameter $\varepsilon$ as a function of the hidden photon mass, including those from SENSEI~\cite{PhysRevLett.134.011804}, SuperCDMS~\cite{PhysRevD.111.012006}, XENONnT~\cite{aprile2024searchlightdarkmatter}, 
      and XENON10 data reinterpreted in Ref.~\cite{EssigHP}; above $20\,\rm{MeV}/c^2$ DAMIC-M sensitivity drops since the selection is limited to $\le5\,e^-$ events.  
    (\textbf{right}) Upper limits on DM-nucleon cross section $\rm{\bar{\sigma}_n}$ through the Migdal effect for a heavy mediator. Limits from  DarkSide-50~\cite{PhysRevLett.130.101001}, 
    PandaX-4T~\cite{PhysRevLett.131.191002}
    SENSEI~\cite{PhysRevLett.134.011804}, 
    SuperCDMS~\cite{PhysRevD.107.112013}, and
    XENON1T~\cite{PhysRevLett.123.241803} using the Migdal effect are also shown, together with limits on DM-nucleon elastic scattering from CRESST~\cite{PhysRevD.100.102002,PhysRevD.110.083038} and SuperCDMS~\cite{PhysRevLett.127.061801}. In the shaded region, the search is no longer sensitive due to Earth-shielding at large cross-sections. The reported Migdal constraint close to 1 MeV may be altered by this effect and we leave a detailed study to future work.
    }
\label{fig:upperlimits1}
 \end{figure*}
\textit{Cross-checks.---}
An independent analysis, utilizing a different calibration and more basic but stringent selection criteria, yields consistent results. This analysis, which was defined before unblinding, provides a cross-check of the selection procedure and gives us confidence in the result obtained.

A check of the background estimation is obtained from the data, by searching for patterns of consecutive pixels in the vertical or diagonal direction, where no ionization events are expected.  We find 139 (130) vertical (diagonal) $\{11\}$ patterns, consistent with the expectation of $B^\mathrm{rc}_{p}=$ 141.4, and zero candidates for all other patterns. 

The uncertainty on $S_p$ and $B_p$ may affect the exclusion limits. For $S_p$, we vary the charge diffusion according to the uncertainty of $\mathrm{\sigma^2_{xy}}$ versus interaction depth; we also vary $\sigma_{ch}$, which is obtained from the data, within its statistical uncertainty. The corresponding changes in  $P_{q_e\rightarrow p}$ and pattern selection result in $<5\%$ difference in the exclusion limits. We find the effect of the statistical uncertainty on $B^\mathrm{rc}_p$ to be negligible. We evaluate the impact of the background from radioactive decays by neglecting it in the likelihood ($B^{\mathrm{rad}}_p=0$), which results in $p_0=0.08$ (0.02) for an ultra-light (heavy) mediator and at most 30\% weaker DM-interaction constraints. 

Significant progress has been made on the theoretical calculation of the DM-$e^-$ interaction rate. More accurate models than the original $\texttt{QEDark}$~\cite{Essig:2015cda} are now available, which properly treat in-medium screening effects, including $\texttt{DarkELF}$~\cite{Knapen_2022}, $\texttt{EXCEED-DM}$~\cite{Griffin_2021,exceedDMpublished} and $\texttt{QCDark}$~\cite{PhysRevD.109.115008}. The latter two feature the most complete calculations including the all-electron effects, and are in good agreement~\cite{PhysRevD.109.115008}. For our limits we use $\texttt{QCDark}$, which is computationally fast. The ratio of $\texttt{QCDark}$ and $\texttt{EXCEED-DM}$-derived limits is $\sim1.1$ for $m_\chi>3$\,MeV/c$^2$, and decreases down to $\sim0.5$ at $m_\chi=1$\,MeV/c$^2$~\footnote[77]{The DAMIC-M 90\% C.L. exclusion limits obtained for several DM candidates and different calculations of the DM-$e^{-}$ interaction rate can be found at \url{https://github.com/DAMIC-M/LBC_2025_HSBenchmark_Limits}.}.

\textit{Upper limits on other DM candidates.---}
Other processes beyond the DM-$e^-$ scattering of Eq.~\eqref{eq:dR_dEe} can result in a charge ionization signal, and can thus be constrained by this search. One is the absorption of a massive vector boson DM, the hidden photon~\cite{Fabbrichesi:2020wbt}, by electrons in the silicon bulk of the CCDs. Another is DM scattering off nuclei, resulting in the generation of a secondary electron via Migdal effect~\cite{Vergados:2005dpd,Moustakidis:2005gx,Bernabei:2007jz,Ibe:2017yqa}.  The corresponding 90\% C.L. exclusion upper limits are shown in  Fig.~\ref{fig:upperlimits1}. 
These are the most stringent constraints to date on hidden photon DM with mass between 2.5 and 24\,eV/c$^2$ and on DM-nucleus scattering for DM masses between 1 and 35\,MeV/c$^2$. We note, however, that the Migdal effect has not been experimentally calibrated yet, and limits from DAMIC-M and other experiments~\cite{PhysRevLett.130.101001,PhysRevLett.131.191002,PhysRevLett.134.011804,PhysRevD.107.112013,PhysRevLett.123.241803} are based on theoretical calculations.

To establish these limits we followed the same procedure used for DM-$e^-$ scattering, with signal rates in Eq.~\eqref{eq:dR_dEe} modified according to the process. 
Specifically, we calculate the hidden photon absorption rates following Ref.~\cite{PhysRevD.105.015014}. As in Ref.~\cite{PhysRevLett.118.141803}, we use the complex index of refraction in silicon from Ref.~\cite{Edwards1997547}, extrapolated to the CCD operating temperature of 130 K with the empirical parametrization from Ref.~\cite{RAJKANAN1979793}. 
The Migdal rates are calculated with the method of Ref.~\cite{Berghaus2023Migdal} using the precomputed dielectric lookup table from Ref.~\cite{Enkovaara_2010, PhysRevD.105.015014}. 

DAMIC-M limits are made available online~\footnotemark[77], including the Migdal limit for a light mediator, not shown in Fig.~\ref{fig:upperlimits1}. 

\end{document}

%% file: 0_authorlist.tex

\author{K.\,Aggarwal}
\affiliation{Center for Experimental Nuclear Physics and Astrophysics, University of Washington, Seattle, WA, United States}

\author{I.\,Arnquist}
\affiliation{Pacific Northwest National Laboratory (PNNL), Richland, WA, United States} 

\author{N.\,Avalos}
\affiliation{Centro At\'{o}mico Bariloche and Instituto Balseiro, Comisi\'{o}n Nacional de Energ\'{i}a At\'{o}mica (CNEA), Consejo Nacional de Investigaciones Cient\'{i}ficas y T\'{e}cnicas (CONICET), Universidad Nacional de Cuyo (UNCUYO), San Carlos de Bariloche, Argentina}

\author{X.\,Bertou}
\affiliation{CNRS/IN2P3, IJCLab, Universit\'{e} Paris-Saclay, Orsay, France}
\affiliation{Laboratoire de physique nucl\'{e}aire et des hautes \'{e}nergies (LPNHE), Sorbonne Universit\'{e}, Universit\'{e} Paris Cit\'{e}, CNRS/IN2P3, Paris, France}

\author{N.\,Castell\'{o}-Mor}
\affiliation{Instituto de F\'{i}sica de Cantabria (IFCA), CSIC - Universidad de Cantabria, Santander, Spain}

\author{A.E.\,Chavarria}
\affiliation{Center for Experimental Nuclear Physics and Astrophysics, University of Washington, Seattle, WA, United States}

\author{J.\,Cuevas-Zepeda}
\affiliation{Kavli Institute for Cosmological Physics and The Enrico Fermi Institute, The University of Chicago, Chicago, IL, United States}

\author{A.\,Dastgheibi-Fard}
\affiliation{LPSC LSM, CNRS/IN2P3, Universit\'{e} Grenoble-Alpes, Grenoble, France}

\author{C.\,De Dominicis}
\affiliation{Laboratoire de physique nucl\'{e}aire et des hautes \'{e}nergies (LPNHE), Sorbonne Universit\'{e}, Universit\'{e} Paris Cit\'{e}, CNRS/IN2P3, Paris, France}

\author{O.\,Deligny}
\affiliation{CNRS/IN2P3, IJCLab, Universit\'{e} Paris-Saclay, Orsay, France}

\author{J.\,Duarte-Campderros}
\affiliation{Instituto de F\'{i}sica de Cantabria (IFCA), CSIC - Universidad de Cantabria, Santander, Spain}

\author{E.\,Estrada}
\affiliation{Centro At\'{o}mico Bariloche and Instituto Balseiro, Comisi\'{o}n Nacional de Energ\'{i}a At\'{o}mica (CNEA), Consejo Nacional de Investigaciones Cient\'{i}ficas y T\'{e}cnicas (CONICET), Universidad Nacional de Cuyo (UNCUYO), San Carlos de Bariloche, Argentina}

\author{R.\,Ga\"{i}or}
\affiliation{Laboratoire de physique nucl\'{e}aire et des hautes \'{e}nergies (LPNHE), Sorbonne Universit\'{e}, Universit\'{e} Paris Cit\'{e}, CNRS/IN2P3, Paris, France}

\author{E.-L.~Gkougkousis}
\affiliation{Universit\"{a}t Z\"{u}rich Physik Institut, Z\"{u}rich, Switzerland}

\author{T.\,Hossbach}
\affiliation{Pacific Northwest National Laboratory (PNNL), Richland, WA, United States} 

\author{L.\,Iddir}
\affiliation{Laboratoire de physique nucl\'{e}aire et des hautes \'{e}nergies (LPNHE), Sorbonne Universit\'{e}, Universit\'{e} Paris Cit\'{e}, CNRS/IN2P3, Paris, France}

\author{B.~J.~Kavanagh}
\affiliation{Instituto de F\'{i}sica de Cantabria (IFCA), CSIC - Universidad de Cantabria, Santander, Spain}

\author{B.\,Kilminster}
\affiliation{Universit\"{a}t Z\"{u}rich Physik Institut, Z\"{u}rich, Switzerland}

\author{A.\,Lantero-Barreda}
\affiliation{Instituto de F\'{i}sica de Cantabria (IFCA), CSIC - Universidad de Cantabria, Santander, Spain}

\author{I.\,Lawson}
\affiliation{SNOLAB, Lively, ON, Canada }

\author{A.\,Letessier-Selvon}
\affiliation{Laboratoire de physique nucl\'{e}aire et des hautes \'{e}nergies (LPNHE), Sorbonne Universit\'{e}, Universit\'{e} Paris Cit\'{e}, CNRS/IN2P3, Paris, France}

\author{H.\,Lin}
\affiliation{Center for Experimental Nuclear Physics and Astrophysics, University of Washington, Seattle, WA, United States}

\author{P.\,Loaiza}
\affiliation{CNRS/IN2P3, IJCLab, Universit\'{e} Paris-Saclay, Orsay, France}

\author{A.\,Lopez-Virto}
\affiliation{Instituto de F\'{i}sica de Cantabria (IFCA), CSIC - Universidad de Cantabria, Santander, Spain}

\author{R.\,Lou}
\affiliation{Kavli Institute for Cosmological Physics and The Enrico Fermi Institute, The University of Chicago, Chicago, IL, United States}

\author{K.~J.\,McGuire}
\affiliation{Center for Experimental Nuclear Physics and Astrophysics, University of Washington, Seattle, WA, United States}

\author{S.\,Munagavalasa}
\affiliation{Kavli Institute for Cosmological Physics and The Enrico Fermi Institute, The University of Chicago, Chicago, IL, United States}

\author{J.\,Noonan}
\affiliation{Kavli Institute for Cosmological Physics and The Enrico Fermi Institute, The University of Chicago, Chicago, IL, United States}

\author{D.\,Norcini}
\affiliation{Department of Physics and Astronomy, Johns Hopkins University, Baltimore, MD, United States}

\author{S.\,Paul}
\affiliation{Kavli Institute for Cosmological Physics and The Enrico Fermi Institute, The University of Chicago, Chicago, IL, United States}

\author{P.\,Privitera}
\affiliation{Kavli Institute for Cosmological Physics and The Enrico Fermi Institute, The University of Chicago, Chicago, IL, United States}
\affiliation{Laboratoire de physique nucl\'{e}aire et des hautes \'{e}nergies (LPNHE), Sorbonne Universit\'{e}, Universit\'{e} Paris Cit\'{e}, CNRS/IN2P3, Paris, France}
 \affiliation{Laboratoire de Physique, \'{E}cole Normale Sup\'{e}rieure, Sorbonne Universit\'{e}, Universit\'{e} Paris Cit\'{e}, CNRS/IN2P3, Paris, France}

\author{P.\,Robmann}
\affiliation{Universit\"{a}t Z\"{u}rich Physik Institut, Z\"{u}rich, Switzerland}

\author{B.\,Roach}
\affiliation{Kavli Institute for Cosmological Physics and The Enrico Fermi Institute, The University of Chicago, Chicago, IL, United States}

\author{M.\,Settimo}
\affiliation{SUBATECH, Nantes Universit\'{e}, IMT Atlantique, CNRS/IN2P3, Nantes, France}

\author{R.\,Smida}
\affiliation{Kavli Institute for Cosmological Physics and The Enrico Fermi Institute, The University of Chicago, Chicago, IL, United States}

\author{M.\,Traina}
\affiliation{Instituto de F\'{i}sica de Cantabria (IFCA), CSIC - Universidad de Cantabria, Santander, Spain}
\affiliation{Center for Experimental Nuclear Physics and Astrophysics, University of Washington, Seattle, WA, United States}

\author{R.\,Vilar}
\affiliation{Instituto de F\'{i}sica de Cantabria (IFCA), CSIC - Universidad de Cantabria, Santander, Spain}

\author{R.\,Yajur}
\affiliation{Kavli Institute for Cosmological Physics and The Enrico Fermi Institute, The University of Chicago, Chicago, IL, United States}

\author{D.\,Venegas-Vargas}
\affiliation{Department of Physics and Astronomy, Johns Hopkins University, Baltimore, MD, United States}

\author{C.\,Zhu}
\affiliation{Department of Physics and Astronomy, Johns Hopkins University, Baltimore, MD, United States}

\author{Y.\,Zhu}
\affiliation{Laboratoire de physique nucl\'{e}aire et des hautes \'{e}nergies (LPNHE), Sorbonne Universit\'{e}, Universit\'{e} Paris Cit\'{e}, CNRS/IN2P3, Paris, France}

\collaboration{DAMIC-M Collaboration}